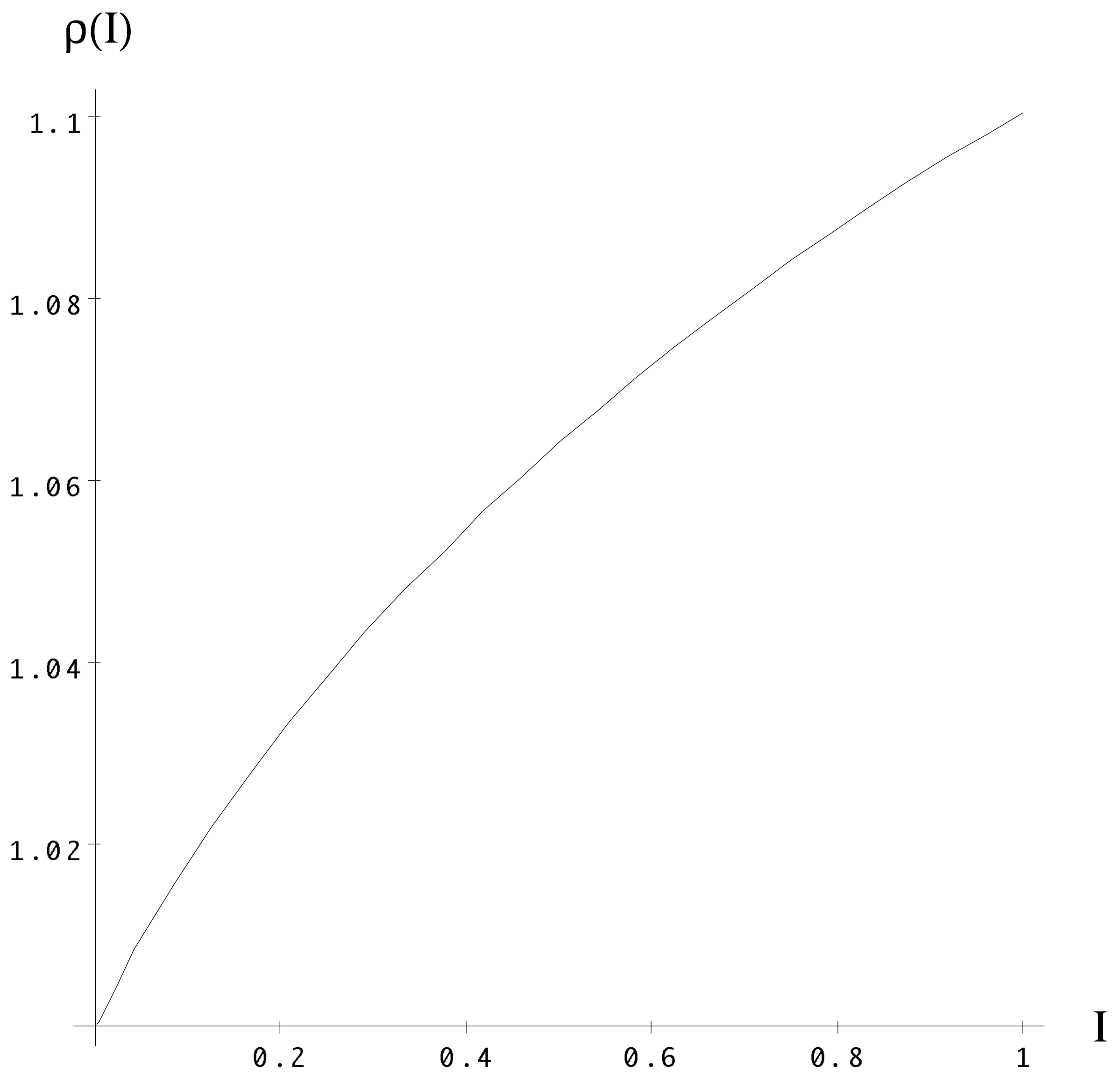

Figure 1. Periodic Orbits of SAM: Nunes, Casasayas, Tufillaro.

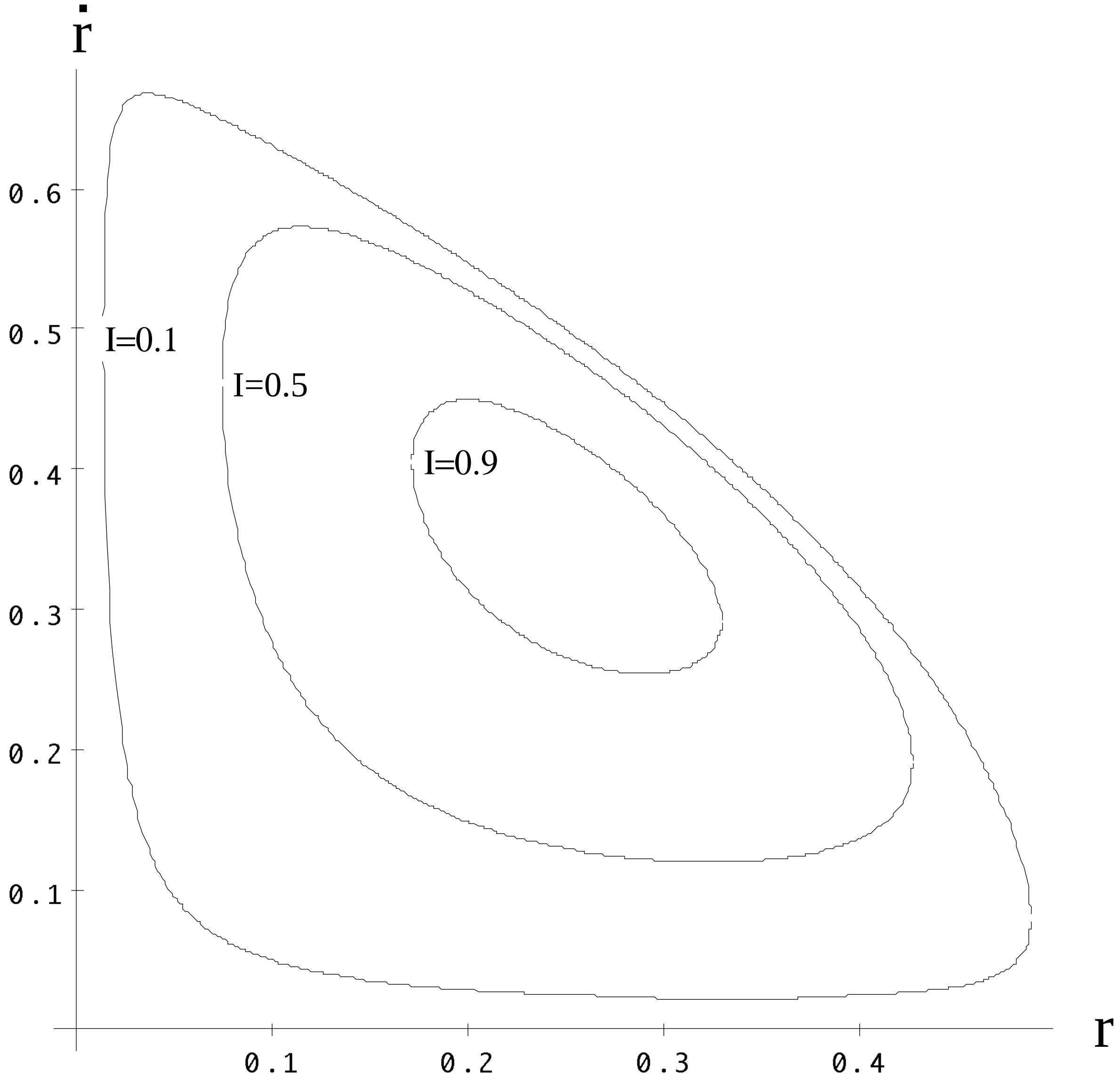

Figure 2. Periodic Orbits of SAM: Nunes, Casasayas, Tufillaro

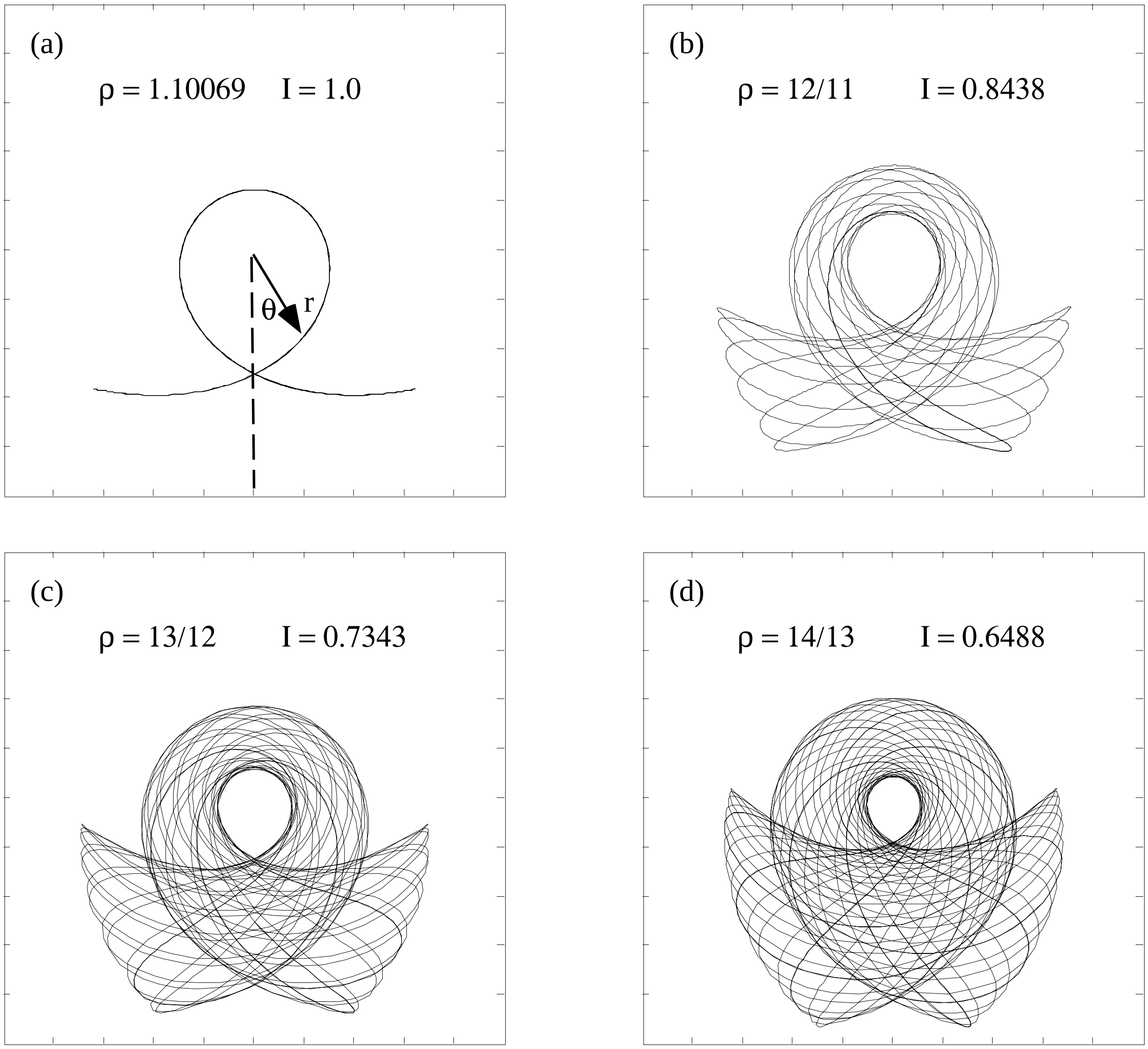

Figure 3. Periodic Orbits of SAM: Nunes, Casasayas, Tufillaro.

# Periodic orbits of the integrable swinging Atwood's machine

Ana Nunes

*Departamento de Fisica*

*Universidade de Lisboa*

*Campo Grande, C1, Piso 4*

*1700 Lisboa, Portugal*

Josefina Casasayas

*Departament de Matemática Aplicada i Analisi*

*Universitat de Barcelona, Gran Via 585*

*08071 Barcelona, Spain*

Nicholas Tufillaro*

*Center for Nonlinear Studies and T13, MS-B258*

*Los Alamos National Laboratories, Los Alamos, NM 87545 USA*

## Abstract

We identify *all* the periodic orbits of the integrable swinging Atwood's machine by calculating the rotation number of each orbit on its invariant tori in phase space, and also providing explicit formulas for the initial conditions needed to generate each orbit.

## I. INTRODUCTION.

Integrable Hamiltonian systems typically display an infinity of distinct periodic and quasiperiodic orbits. The bounded Kepler problem often studied in classical mechanics is not typical in this respect since it only exhibits periodic orbits of a simple type. This is because the Kepler problem has a third invariant, the Runge-Lenz vector [1], in addition to the energy and angular momentum. This third invariant forces all the bounded motions to be periodic. A somewhat more typical example of an integrable two-degree of freedom Hamiltonian system is the swinging Atwood's machine when the mass ratio ($\mu = M/m$) of the non-swinging to swinging mass is equal to three [2,3].

Numerical studies of the integrable swinging Atwood's machine exhibit a plethora of distinct periodic and quasiperiodic orbits, and when viewed in configuration space it appears to be a difficult task to organize and classify all these different types of orbits.

The modern view of classical mechanics emphasizes the "geometry of phase space," and with this geometric perspective the problem of the classification of orbits in integrable Hamiltonian systems becomes elementary. A basic result of the geometric approach to classical mechanics states that the topology of the invariant manifold $M$ of integrable Hamiltonian systems must be an $n$-dimensional torus [4]. In particular, for a two-degree of freedom Hamiltonian system with just two invariants, each orbit in the four-dimensional phase space must be confined to a two-dimensional torus, a hollow donut. Furthermore, all orbits which arise from flows (vector fields) on a two-dimensional torus must be either a periodic or quasiperiodic winding of the torus. In the latter case the quasiperiodic orbit will densely cover the whole torus.

The "natural" way then to classify the orbits is to calculate rotation number about the torus: the ratio of winding about each generator of the torus. For quasiperiodic orbits this rotation number is irrational, and it is a single number that uniquely identifies the orbit. In the case of a periodic orbit, this number is rational and it uniquely identifies a family of periodic orbits all of the same type or shape when viewed in configuration space. Thus the rotation number, which uniquely indexes each torus (and hence each type of orbit) in the phase space of an integrable Hamiltonian system, provides the most natural solution to the problem of identifying the orbits of a two-degree of freedom integrable Hamiltonian system.

In this paper we use some classical results of classical mechanics (Hamilton-Jacobi theory) along with the modern perspective of classical mechanics (the geometry of phase space) in order to identify and classify *all* the periodic orbits of the integrable swinging Atwood's machine. It is hoped that this elementary but physically realistic example might find use in an advanced classical mechanics course which attempts to introduce the flavor of a modern approach to classical mechanics.

## II. THE ORBIT EQUATION.

Taking polar coordinates in the vertical plane where the weight of swinging mass $m$ moves, the total energy of the system is

$$\begin{aligned} E &= T + V \\ &= \frac{M+m}{2}\dot{r}^2 + \frac{1}{2}mr^2\dot{\theta}^2 + gr(M - m\cos\theta). \end{aligned} \tag{1}$$

The conjugate momenta $p_r$ and $p_\theta$ are

$$p_r = (M+m)\dot{r}, \tag{2a}$$

$$p_\theta = mr^2\dot{\theta}. \tag{2b}$$

So the Hamiltonian of the problem is given by

$$\begin{aligned} &H(r,\theta,p_r,p_\theta) = \\ &\qquad \frac{1}{2}\left(\frac{p_r^2}{M+m} + \frac{p_\theta^2}{mr^2}\right) + gr(M - m\cos\theta). \end{aligned} \tag{3}$$

As shown in [3], it is convenient to change to another set of canonical coordinates where the equations of motion become simpler, especially if we want to do Hamilton-Jacobi theory. Let the coordinate transformation to $(\xi, \eta)$ be defined by

$$\xi = \sqrt{r(1+\sin(\theta/2))}, \tag{4a}$$

$$\eta = \pm\sqrt{r(1-\sin(\theta/2))}. \tag{4b}$$

The inverse change of coordinates is

$$r = (\xi^2+\eta^2)/2, \tag{5a}$$

$$\theta = 2\arctan\left[(\xi^2-\eta^2)/2\xi\eta\right], \tag{5b}$$

and the generating function $F_2(r,\theta,p_\xi,p_\eta)$ which determines the extension of this change of position coordinates to a canonical transformation in phase space will be

$$F_2(r,\theta,p_\xi,p_\eta) = \sqrt{r(1+\sin(\theta/2))}\ p_\xi + \sqrt{r(1-\sin(\theta/2))}\ p_\eta. \tag{6}$$

Hence,

$$\begin{aligned} p_r &= \frac{\partial F_2}{\partial r} \\ &= \frac{\sqrt{1+\sin(\theta/2)}}{2\sqrt{r}} p_\xi + \frac{\sqrt{1-\sin(\theta/2)}}{2\sqrt{r}} p_\eta \\ &= \frac{\xi p_\xi + \eta p_\eta}{\xi^2+\eta^2}, \end{aligned} \tag{7a}$$

$$\begin{aligned} p_\theta &= \frac{\partial F_2}{\partial \theta} \\ &= \frac{\sqrt{r}\cos(\theta/2)}{4\sqrt{1+\sin(\theta/2)}} p_\xi - \frac{\sqrt{r}\cos(\theta/2)}{4\sqrt{1-\sin(\theta/2)}} p_\eta \\ &= \frac{\eta p_\xi - \xi p_\eta}{4}. \end{aligned} \tag{7b}$$

Since we know (see [5]) that the system is integrable only when $M/m = 3$, we set $M = 3$ and $m = 1$ throughout this paper. For these values of the parameters, eq. (3) becomes, after introducing eqs. (5) and eqs. (7),

$$H(\xi,\eta,p_\xi,p_\eta) = \frac{1}{\xi^2+\eta^2}\left[\frac{1}{8}(p_\xi^2+p_\eta^2) + 2g(\xi^4+\eta^4)\right], \tag{8}$$

the expression of the Hamiltonian of the system in the new set of coordinates.

We recall that the idea behind Hamilton-Jacobi theory is to find a generating function $S(\xi, \eta, \beta_1, \beta_2, t)$ for a canonical change from $(\xi, \eta, p_\xi, p_\eta)$ to a new set of variables $(\alpha_1, \alpha_2, \beta_1, \beta_2)$ which are constant along the motion, i.e., such that the Hamiltonian expressed in terms of the new variables is identically zero. The function $S$ with this property satisfies the first order partial differential equation

$$H\left(\xi, \eta, \frac{\partial S}{\partial \xi}, \frac{\partial S}{\partial \eta}\right) + \frac{\partial S}{\partial t} = 0, \tag{9}$$

and the constants $\beta_1, \beta_2$ will be the two independent constants which will come out of the integration of eq. (9). Therefore, by construction, solving eq. (9) is equivalent to solving the equations of motion.

As shown in [6], eq. (9) for Hamiltonian eq. (8) separates and yields a solution in terms of quadratures, i.e., the problem reduces to that of computing certain integrals. More precisely, let us consider eq. (9), where $H$ is the Hamiltonian given by eq. (8), and try a solution of the form

$$S(\xi, \eta, t) = S_1(\xi) + S_2(\eta) + S_3(t). \tag{10}$$

Then, eq. (9) becomes

$$\frac{1}{\xi^2 + \eta^2}\left[\frac{1}{8}\left(\frac{dS_1}{d\xi}^2 + \frac{dS_2}{d\eta}^2\right) + 2g(\xi^4 + \eta^4)\right] = -\frac{dS_3}{dt}. \tag{11}$$

Clearly, each member must be equal to a constant, $\beta_1$, and moreover, since by definition

$$p_\xi = \frac{\partial S}{\partial \xi}, \tag{12a}$$

$$p_\eta = \frac{\partial S}{\partial \eta}, \tag{12b}$$

this constant must be the total energy of the system, i.e.,

$$\beta_1 = E, \tag{13}$$

and so

$$S_3(t) = -Et. \tag{14}$$

Equation (9) then reduces to another separable equation in $\xi$ and $\eta$,

$$\frac{1}{8}\left(\frac{dS_1}{d\xi}\right)^2 + 2g\xi^4 - E\xi^2 = \\ -\frac{1}{8}\left(\frac{dS_2}{d\eta}\right)^2 - 2g\eta^4 + E\eta^2. \tag{15}$$

Again, both members must be equal to the second integration constant $\beta_2$, which we shall name $I/8$. Therefore, denoting by $s(x, E, I)$,

$$s(x, E, I) = \int_k^x \sqrt{I - 16gy^4 + 8Ey^2}dy, \tag{16}$$

where $k$ is some constant to be fixed later, the solution $S(\xi, \eta, E, I, t)$ is, up to two independent additive constants,

$$S(\xi, \eta, E, I, t) = s(\xi, E, I) + s(\eta, E, -I) - Et. \tag{17}$$

The equations of the canonical transformation whose generating function is $S$ which will give the complete solution of the problem are

$$\alpha_1 = \frac{\partial S}{\partial E} = \frac{\partial s}{\partial E}(\xi, E, I) + \frac{\partial s}{\partial E}(\eta, E, -I) - t, \tag{18a}$$

$$\alpha_2 = \frac{\partial s}{\partial I}(\xi, E, I) + \frac{\partial s}{\partial I}(\eta, E, -I), \tag{18b}$$

$$p_\xi = \frac{\partial S}{\partial \xi} = \frac{\partial s}{\partial \xi}(\xi, E, I), \tag{18c}$$

$$p_\eta = \frac{\partial s}{\partial \eta} = \frac{\partial s}{\partial \eta}(\eta, E, -I). \tag{18d}$$

Note that the two independent additive constants are now included in eqs. (18a) and (18b). Equation (18b) is the orbit equation, which gives a relation between the position coordinates $\xi$ and $\eta$ in terms of the constants $E$, $I$, and $\alpha_2$. Adding the information contained in eq. (18a), that contains time explicitly, we shall obtain the trajectories equations. It is therefore clear that the other constant $\alpha_1$ plays the role of the initial time. The last two equations of (18) complete the description of the motion in phase space.

Let us now focus our attention on the orbit equation, to obtain an explicit expression in terms of known functions. First, we note that, as pointed out in [7], we may always choose appropriate time and length units to set $E = g = 1$. From now on, we shall assume that we are working in these units. In particular, the parameter $I$ will always take values in the interval $[-1, 1]$. Then, the orbit equation becomes

$$\begin{aligned}\alpha_2 &= \frac{1}{2}\int_{k_1}^{\xi} \frac{dy}{\sqrt{I - 16y^4 + 8y^2}} - \frac{1}{2}\int_{k_2}^{\eta} \frac{dy}{\sqrt{-I - 16y^4 + 8y^2}} \\ &= \frac{1}{8}\Xi(\xi, I) + \frac{1}{8}\Theta(\eta, I). \end{aligned} \tag{19}$$

Fixing the constants $k_1$ and $k_2$ at appropriate values and assuming for now that $I \geq 0$, we obtain

$$\Xi(\xi, I) = Y_1(\xi, I), \tag{20a}$$

$$\Theta(\eta, I) = Y_2(\eta, I), \tag{20b}$$

where

$$Y_1(x, I) = \int_0^x \frac{dy}{\sqrt{(a^2 - y^2)(b^2 + y^2)}}, \tag{21a}$$

$$Y_2(x, I) = \int_x^c \frac{dy}{\sqrt{(c^2 - y^2)(y^2 - d^2)}}, \tag{21b}$$

and

$$\begin{aligned} a^2 &= \frac{1 + \sqrt{1 + I}}{4}, \quad & b^2 &= \frac{-1 + \sqrt{1 + I}}{4}, \\ c^2 &= \frac{1 - \sqrt{1 - I}}{4}, \quad & d^2 &= \frac{1 + \sqrt{1 - I}}{4}. \end{aligned} \tag{22}$$

On the other hand, we also have, for $I < 0$,

$$\Xi(\xi, I) = -Y_2(\xi, -I), \tag{23a}$$

$$\Theta(\eta, I) = -Y_1(\eta, -I). \tag{23b}$$

Therefore, the orbit equation may always be expressed in terms of the functions $Y_1(x, I)$ and $Y_2(x, I)$, $I \geq 0$. Now, both $Y_1$ and $Y_2$ are elliptic integrals of the first kind (see, for instance, [8]). More precisely, we have, denoting as usual by $F(\varphi|m)$ the elliptic integral

$$F(\varphi|m) = \int_0^{\varphi} (1 - m\sin^2\phi)^{-1/2} d\phi$$
$$= \int_0^{z=\sin\varphi} [(1-y^2)(1-my^2)]^{-1/2} dy, \tag{24}$$

$$Y_1(x, I) = \sqrt{\frac{2}{\sqrt{1+I}}} F(\varphi_1|m_1), \tag{25a}$$
$$Y_2(x, I) = \sqrt{\frac{4}{1+\sqrt{1-I}}} F(\varphi_2|m_2) \tag{25b}$$

with

$$m_1(I) = \frac{1+\sqrt{1+I}}{2\sqrt{1+I}}, \tag{26a}$$
$$m_2(I) = \frac{2\sqrt{1-I}}{1+\sqrt{1-I}}, \tag{26b}$$

and

$$\sin\varphi_1(x, I) = \sqrt{\frac{8x^2\sqrt{1+I}}{4x^2(1+\sqrt{1+I})+I}}, \tag{27a}$$
$$\sin\varphi_2(x, I) = \sqrt{\frac{1-4x^2+\sqrt{1-I}}{2\sqrt{1-I}}}. \tag{27b}$$

Hence, for $I \geq 0$ the orbit equation may be written

$$\alpha_2 = \sqrt{\frac{1}{\sqrt{2+2I}}} F(\varphi_1(\xi, I)|m_1(I))$$
$$+ \sqrt{\frac{\sqrt{2}}{1+\sqrt{1-I}}} F(\varphi_2(\eta, I)|m_2(I)), \tag{28}$$

where $\varphi_1(\xi, I)$, $m_1(I)$, $\varphi_2(\eta, I)$, and $m_2(I)$ are given by eqs. (26) and (27). For $I < 0$, by eq. (23), the orbit equation is obtained from eq. (28) by interchanging $\xi$ and $\eta$.

Using eq. (28) and a symbolic math program, and going back to the original coordinates $r, \theta$, it is easy to obtain plots of the orbits of the system in the plane.

## III. LOW-ORDER PERIODIC ORBITS (TORUS KNOTS)

The system we are dealing with is non-degenerate and so we expect to find periodic and quasi-periodic orbits, according to the value of $I$ that we pick in a certain energy level. Since we are studying an integrable Hamiltonian system, all these orbits, both the periodic and the quasi-periodic will lie on tori in the phase space, which form the closure of the quasi-periodic orbits in one case, and which will be foliated by the periodic orbits in the other case. This determines the knot type of the periodic orbits: all of them must be torus knots. A torus knot is characterized by two integers, $n, m$, that measure the number of times the curve winds around each generator of the torus. The rational rotation number $\rho$ associated to each of the tori foliated by periodic orbits will measure the ratio $n/m$. Therefore, if we want to know which torus knots are realized as orbits of our system, or if we want to know how to pick the "simplest" torus knots, i.e., those for which both $n$ and $m$ are small integers, we need information on the dependence of the rotation number on the parameter $I$. The range of the function $\rho(I)$ will tell us which are the low-order resonances and hence which are the low-order torus knots, and the form of the function itself will tell us where to look for these knots, i.e., which approximate value for $I$ we should take. The appropriate variables to study this question are action-angle variables.

Recall that, in general, the action variable $J_i$ is given by $\oint p_i dq_i$, where the integral is taken along a period of the projection of the orbit on the $q_i, p_i$ plane. The angles are the canonical conjugate of the actions, and the Hamiltonian in this set of variables depends only on the $J_i$. The angles change with time with constant frequencies, given by the partial derivatives of $H$ with respect to the $J_i$. Hence, in our case, the rotation number associated to the motion on the two-dimensional tori that foliate the energy level $E = 1$ will be given by

$$\rho(I) = \frac{\partial H/\partial J_1}{\partial H/\partial J_2}, \tag{29}$$

where the derivatives must be evaluated along $H(J_1, J_2) = 1$. Now, from equation (6.3) and (6.4) we have

$$\begin{aligned} J_1(E, I) &= \oint p_\xi d\xi = \oint \frac{\partial S}{\partial \xi} d\xi \\ &= \oint \sqrt{I + 8E\xi^2 - 16\xi^4} d\xi = j(E, I) \end{aligned} \tag{30a}$$

$$J_2(E,I) = \oint p_\eta d\eta = \oint \frac{\partial S}{\partial \eta} d\eta$$
$$= \oint \sqrt{-I + 8E\eta^2 - 16\eta^4} d\eta = j(E, -I), \tag{30b}$$

where we have recovered the dependence on $E$ since it is essential for the computation of eq. (29). From the identity $E(J_1(E,I), J_2(E,I)) = E$ we get

$$\rho(I) = \frac{\partial E/\partial J_1}{\partial E/\partial J_2} = -\frac{\frac{\partial J_2}{\partial I}(1,I)}{\frac{\partial J_1}{\partial I}(1,I)} = \frac{\frac{\partial j}{\partial I}(1,-I)}{\frac{\partial j}{\partial I}(1,I)}. \tag{31}$$

But

$$\frac{\partial j}{\partial I}(1,I) = \frac{1}{2} \oint \frac{dy}{\sqrt{I + 8y^2 - 16y^4}} \tag{32}$$

and so, for $I \geq 0$,

$$\rho(I) = \frac{\oint \frac{dy}{\sqrt{(c^2 - y^2)(y^2 - d^2)}}}{\oint \frac{dy}{\sqrt{(a^2 - y^2)(y^2 + b^2)}}} \tag{33}$$

where $a$, $b$, $c$ and $d$ are given by eq. (22), while as in eq. (23),

$$\rho(-I) = 1/\rho(I) \tag{34}$$

holds. Hence, it is enough to study the function $\rho$ on the interval $[0,1]$. This last identity arises from the time reversibility of the system since the first integral is invariant under a change in the velocity sign.

As we have already seen, the indefinite integrals correspond to elliptic functions of the first kind. Now, the integrals in eq. (33) are the corresponding complete elliptic integrals of the first kind $K(m)$ (see, for instance, [8]). In terms of these functions, eq. (33) becomes

$$\rho(I) = \sqrt{\frac{2\sqrt{1+I}}{1+\sqrt{1-I}} \frac{K[m_2(I)]}{K[m_1(I)]}}, \tag{35}$$

where $m_1(I)$ and $m_2(I)$ are given by eq. (26). This expression and the range $\rho([0,1])$ can be easily computed using for $K(m)$ the polynomial approximation

$$K(1-m) = (a_0 + a_1 m + a_2 m^2) +$$
$$(b_0 + b_1 m + b_2 m^2)\log(1/m) + \epsilon(m), \tag{36}$$

where $|\epsilon(m)| \leq 3.10^{-5}$ and $a_0 = 1.38629$, $a_1 = 0.11197$, $a_2 = 0.07252$, $b_0 = 0.5$, $b_1 = 0.12134$ and $b_2 = 0.02887$.

Figure 1 is a plot of $\rho(I)$ verses $I$. The graph shows a monotonic function with $\rho(0) = 1$ and $\rho(1) = \pi/(2^{1/4}K[(1+\sqrt{2})/2^{3/2}]) \approx 1.10069$. Thus, no periodic orbit can exist in the integrable swinging Atwood's machine with rotation number greater than about 1.10069. The "simplest" periodic orbits will therefore be of fairly high period and will have rotation numbers like $11/10, 12/11, 13/12$ and so on.

In order to plot some of these periodic orbits it is useful to express the initial conditions as a function of $I$. Then to locate a periodic orbit with a given rational rotation number, all we need to do is calculate the inverse function $\rho^{-1}(n/m) = I$ and then use this value of $I$ to find the initial conditions needed to generated an orbit with rotation number $\rho = n/m$.

To find a formula for the initial conditions we note that at the surface of section $\theta = 0$, the energy $E(r, \theta, \dot{r}, \dot{\theta})$ and first integral $I((r, \theta, \dot{r}, \dot{\theta})$ are with $E = g = 1$ [9]:

$$2\dot{r}^2 + \frac{1}{2}r^2\dot{\theta}^2 + 2r = 1, \tag{37}$$

and,

$$I = 16r^2\dot{r}\dot{\theta}. \tag{38}$$

Eliminating $\dot{\theta}$ between these two invariants we find

$$\dot{r}(r, I) = \pm\frac{1}{\sqrt{2}}\sqrt{\left(\frac{1}{2} - r\right) \pm \sqrt{\left(\frac{1}{2} - r\right)^2 - \frac{I^2}{(16r)^2}}}. \tag{39}$$

All *real* solutions of eq. (39) with $I \in [0, 1]$ represent cross-sections of the invariant tori at $\theta = 0$. A few of these cross-sections are shown in Figure 2. The "first" torus occurs at $I = 0$ and corresponds to the family of "teardrop-heart" ejection-collision orbits previously studied [6]. The "last" torus occurs at $I = 1$ and corresponds with the "loop" periodic orbit shown in Figure 3(a). This last torus is degenerate, and it is the only periodic orbit in the integrable swinging Atwood's machine which may have a non-rational rotation number. All other orbits on the tori between $0 < I < 1$ are either a family of periodic orbits if $\rho(I)$ is rational, or a single quasiperiodic orbit otherwise. Equations (39) and (38) with $r = 0.25$ and $\theta = 0$ provide explicit initial conditions for an orbit with any value of $I$. The "last torus" (loop orbit) has the initial conditions $(r_0, \theta_0, \dot{r}_0, \dot{\theta}_0) \approx (0.25, 0, 0.353553, 2.82843)$.

Figure 3 illustrates a few of the simpler periodic orbits that occur in the integrable swinging Atwood's machine.

## IV. LOOP ORBIT

We conclude by showing that it is possible to write down an explicit exact solution for the loop orbit ($E = g = 1$). In the $(\xi, \eta)$ coordinates, initial conditions for the loop orbit are $\xi = \eta = 1/2$ with $\dot{\xi} = \sqrt{2}/2$ and $\dot{\eta} = 0$. And it is easy to check that $\ddot{\eta} = 0$. So, the solution of the loop orbit in the $(\xi, \eta)$ coordinates is particularly simple, namely: $\xi = \xi(t)$ and $\eta = 1/2$ for all $t$. Now, using eq. (5), we find that on transforming back to the original $(r, \theta)$ coordinates this orbit gets mapped to:

$$r_-(\theta) = \tfrac{1}{4}\sec(\tfrac{\theta}{2})\left[\sec(\tfrac{\theta}{2}) - \tan(\tfrac{\theta}{2})\right] \quad \text{for} \quad 0 \le \theta \le 2\pi, \tag{40a}$$

$$r_+(\theta) = \tfrac{1}{4}\sec(\tfrac{\theta}{2})\left[\sec(\tfrac{\theta}{2}) + \tan(\tfrac{\theta}{2})\right] \quad \text{for} \quad -\theta_0 \le \theta \le \theta_0 \tag{40b}$$

where the range of the solution on the second branch of the orbit (eq. 40b) can be determined by setting the velocities $(\dot{r}, \dot{\theta})$ equal to zero in the energy equation (1),

$$r[3 - \cos(\theta)] = 1 \tag{41}$$

and the second invariant $I(r, \theta, \dot{r}, \dot{\theta})$ [9],

$$16 r^2 \sin\left(\frac{\theta}{2}\right)\cos^2\left(\frac{\theta}{2}\right) = 1, \tag{42}$$

which is equal to one for the loop orbit. Solving for $\theta$ we find:

$$\theta_0 = \arccos(-5 + 4\sqrt{2}) \approx 0.854 \text{ radians}. \tag{43}$$

## REFERENCES

* Internet: nbt@reed.edu.

[1] H. Goldstein, *Classical Mechanics*, 2nd ed. (Addison-Wesley, Reading, MA 1980).

[2] D. J. Griffiths and T. A. Abbott, Comment on "A surprising mechanics demonstration," Am. J. Phys. **60** (10), 951-953 (1992).

[3] N. Tufillaro, Integrable motion of a swinging Atwood's machine, Am. J. Phys. **54**, 142-143 (1986).

[4] M. Tabor, *Chaos and integrability in nonlinear dynamics* (John Wiley and Sons, New York, 1989).

[5] J. Casasayas, A. Nunes, and N. Tufillaro, Swinging Atwood's Machine: integrability and dynamics, J. Phys. France **51**, 1693-1702 (1990).

[6] N. Tufillaro, Teardrop and heart orbits of a swinging Atwood's machine, Am. J. Phys. **62** (3), 231-233 (1994).

[7] N. Tufillaro, Motions of a swinging Atwood's machine, J. Phys. France **46**, 1495-1500 (1985).

[8] M. Abramowitz and I. Stegun, *Handbook of Mathematical Functions* (Dover, New York, 1965). Our notation differs slightly. In particular, $m$ here is $m_1$ in Abramowitz and Stegun.

[9] See Ref. [3]. The value for the invariant $I$ in this paper differs from that in eq. (16) of Ref. [3] (call it $\bar{I}$) by a scale factor: $\bar{I} = I/16$.

## Figure Captions

**Figure 1.** Rotation number $\rho(I)$ as a function of the invariant $I$.

**Figure 2.** Cross sections (at $\theta = 0$) of invariant tori for a few values of $I$. $I = 0$ corresponds to the "teardrop-heart" orbits while $I = 1$ corresponds to the "loop" orbit, a degenerate torus at $r = 0.25$.

**Figure 3.** A few periodic orbits and their rotation numbers with initial conditions $(r_0, \theta_0, \dot{r}_0, \dot{\theta}_0)$: (a) (0.25, 0, 0.353353, 2.82843), (b) (0.25, 0, 0.438269, 1.92533), (c) (0.25, 0, 0.458094, 1.603), (d) (0.25, 0, 0.469171, 1.38283).